\documentclass[%
 reprint,
 amsmath,amssymb,
 aps,
]{revtex4-2}

\usepackage{graphicx}
\usepackage{dcolumn}
\usepackage{bm}

\begin{document}

\preprint{APS/123-QED}

\title{Spin Elasticity: A New Paradigm for Spintronics}

\author{Zhong-Chen Gao }
 \email{Corresponding author: gaozc0129@shnu.edu.cn }
\affiliation{Mathematics and Science College, Shanghai Normal University, Shanghai 200234, China }%

\author{Tianyi Zhang}%
\affiliation{Beijing National Laboratory for Condensed Matter Physics, Institute of Physics, University of Chinese Academy of Sciences, Chinese Academy of Sciences, Beijing 100190, China }%
\author{Feifei Wang }
\affiliation{Key Laboratory of Optoelectronic Material and Device, Department of Physics, Shanghai Normal University, Shanghai 200234, China }%
\author{Jingguo Hu }
\affiliation{School of Physical Science and Technology (School of Integrated Circuits), Yangzhou University, Yangzhou 225002, China }%
\author{Peng Yan }
\affiliation{School of Physics and State Key Laboratory of Electronic Thin Films and Integrated Devices, University of Electronic Science and Technology of China, Chengdu 611731, China }%
\author{Xiufeng Han }
\affiliation{Beijing National Laboratory for Condensed Matter Physics, Institute of Physics, University of Chinese Academy of Sciences, Chinese Academy of Sciences, Beijing 100190, China }%
\affiliation{Center of Materials Science and Optoelectronics Engineering, University of Chinese Academy of Sciences, Beijing 100049, China }%
\affiliation{Songshan Lake Materials Laboratory, Dongguan, Guangdong 523808, China }%

\date{\today}

\begin{abstract}
Elasticity shapes our world. For centuries, it has been regarded as a property exclusive to ordinary matter. Here we uncover its hidden existence in the spin degree of freedom. We introduce \textit{spin elasticity}—a framework linking spin torque to spin morphology. This reveals a topological Hooke's law, uncovers spontaneous oscillations and resonance, and predicts a new class of collective excitations: \textit{spin stress waves}. By establishing a unified $\bm{\tau}-\textbf{D}$ theory bridging classical elasticity and topological spin physics, this work completes the elastic picture and opens a new frontier for spintronics—spin-elastronics.  
\begin{description}

\item[Subject area]
Spintronics, Condensed Matter Physics 
\end{description}
\end{abstract}

\maketitle


\section{\label{sec:level1}Introduction}

Without elasticity, the physical world as we know it would cease to exist. It is elasticity that underpins the stability and structure of all matter around us. More than that, the trajectory of human civilization itself is deeply intertwined with the mastery of this property. From the taut bowstrings of the Pleistocene [1] and the spring-dampened chariots of Tutankhamun [2] to the bronze alloy springs powering water clocks and catapults in Ctesibius’s Alexandria [3], empirically designed elastic components had already become integral to daily life millennia ago. A pivotal turning point came with the formulation of Hooke’s law—Ut tensio, sic vis—in the late 17th century [4], followed by the formal establishment of elasticity theory by Cauchy, Saint-Venant, and others in the 19th century [5,6]. These advances propelled the Industrial Revolution, providing critical theoretical foundations for precision engineering innovations such as the high-pressure steam engine [7] and the Brooklyn bridge [8]. Today, elastic analysis and design permeate every scale of human endeavor—from the soles of our shoes to the aircraft overhead and the Voyager probe now traversing interstellar space. 

At its core, the canonical theory of elasticity describes bodies—whether metals with atomic lattices or rubbers with polymer chains—structured by charge and mass. In this framework, elasticity originates from the spatial arrangement of massive particles governed by charge-mediated intermolecular electromagnetic forces, which share two distinct characteristics: (1) they are attractive at large distances, decaying to zero at infinity; and (2) they are strongly repulsive at short range. Spin, the third fundamental attribute of particles alongside charge and mass, has remained conspicuously absent from this narrative. 

\begin{figure*}
    \centering
    \includegraphics[width=0.9\linewidth]{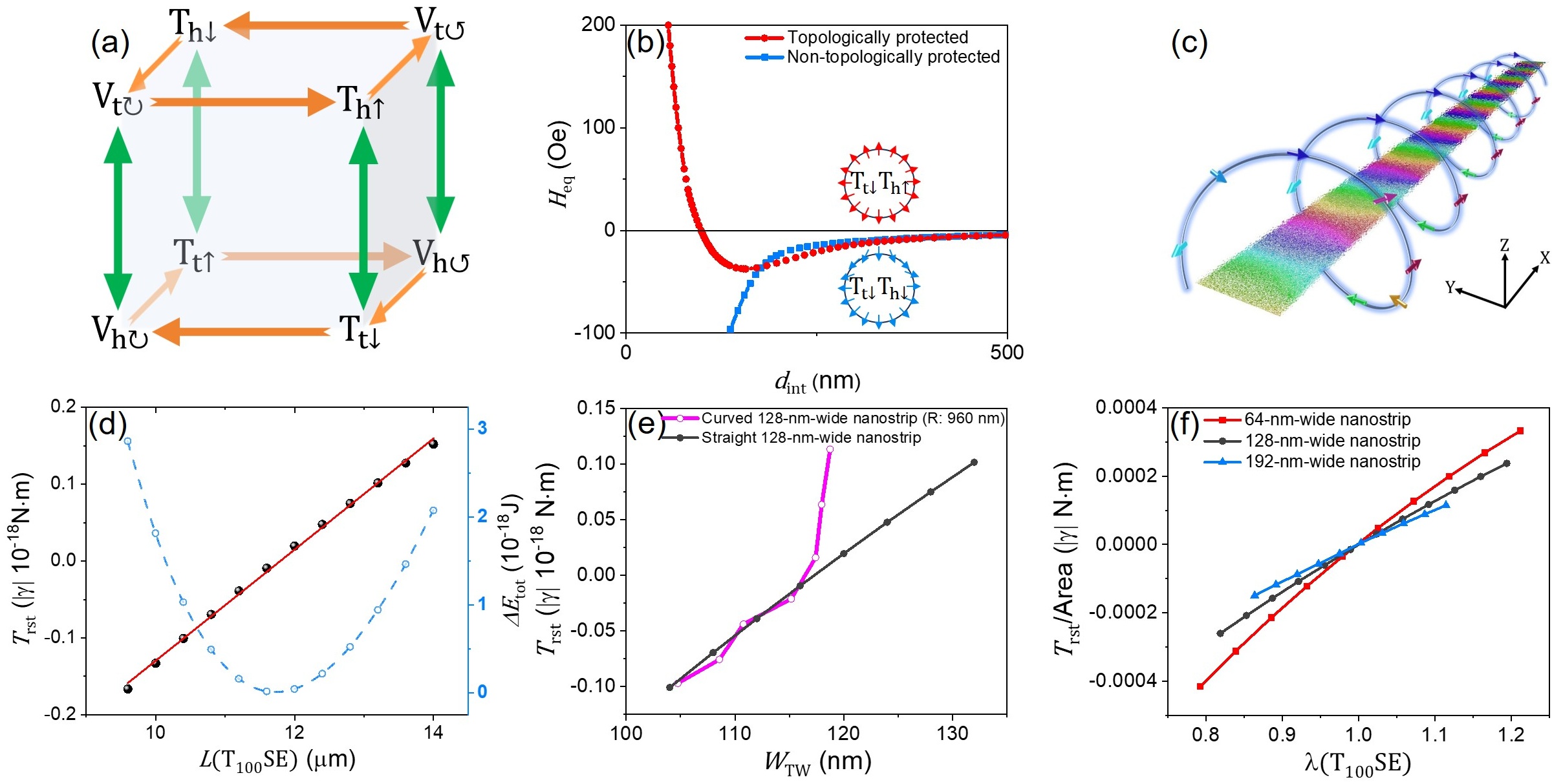}
    \caption{(a) Allowed adjacent pairings between vortex walls and transverse walls (TWs). Arrows indicate permitted arrangements without annihilation. (b) Inter-solitonic interaction between TWs. $H_{\text{eq}}$ is the negative of external field at which the system equilibrates at a given separation $d_{\text{int}}$. (c) Schematic of a $\textnormal{T}_n\textnormal{SE}$ constructed from nontrivial pair $\mathrm{T_{t\downarrow}}\mathrm{T_{h\uparrow}}$. (d) Restoring spin torque $T_{\text{rst}}$ and energy storage $\Delta E_{\text{tot}}$ as functions of length $L(\textnormal{T}_{100}\textnormal{SE})$. (e) $T_{\text{rst}}$ as a function of TW width $W_{\text{TW}}$ in straight and curved 128-nm nanostrips. (f) Width-dependent unit-area $T_{\text{rst}}$ versus elongation ratio $\lambda(\textnormal{T}_{100}\textnormal{SE})$.}
    \label{fig:placeholder}
\end{figure*}

Could elasticity, then, manifest in the spin degree of freedom?

In this Letter, we answer this question in the affirmative. We introduce the concept of \textit{spin elasticity}—an intrinsic mechanism governing the recoverable deformation of spin solitonic textures, which we term \textit{spin elastomers} (SE). Statically, we demonstrate that SE can stretch, compress, and spring back like their mechanical counterparts—but obeying a topological Hooke’s law. Dynamically, we uncover in the massless SE spontaneous oscillations and resonance—another hallmark of elasticity—in direct violation of the Landau–Lifshitz–Gilbert (LLG) equation [9,10], and predict a new class of collective excitations: spin stress waves. We further develop a continuum \textit{$\bm{\tau}-\textbf{D}$} theory that captures the response of internal deformation under load distribution within SE, providing a unified framework that bridges classical elasticity and topological spin physics. These advances reveal a previously unrecognized universality: elasticity operates in both matter and spin spaces, underpinning structural integrity across physical realms.

\section{\label{sec:level2}Model}

We consider a soft ferromagnetic nanostrip governed by the Hamiltonian
\begin{equation}
\mathcal{H} =\int\left[A(\nabla\textbf{m})^2-\frac{1}{2}\mu_0 M_s \textbf{m}\cdot \textbf{H}_d \right]d\bm{r},  
\end{equation}
where $\textbf{m}=\textbf{M}/M_s$ is the normalized magnetization, $A$ the exchange constant, $\mu_0$ the vacuum permeability, $\textbf{H}_d$ the dipolar field. The competition between the exchange and dipolar interactions stabilizes two types of magnetic solitons—vortex and transverse walls (TWs) [11,12]. The allowed adjacent solitonic pairings that are compliant with magnetization homotopy [12] are given in Fig. 1(a). Focusing on the TWs, their interaction diagram (Fig. 1(b)) reveals that for topologically protected pairs, a strong repulsion emerges upon approach, competing with the long-range magnetostatic attraction. Such competition yields an interaction curve akin to that of atomic potentials, setting the stage for spin elasticity. Fig. 1(c) depicts a 1-D SE constructed from the nontrivial $\mathrm{T_{t\downarrow}}\mathrm{T_{h\uparrow}}$ pair (a "$\textnormal{T}\textnormal{SE}$" or "$\textnormal{T}_n\textnormal{SE}$" where $n$ denotes TW number). Detailed magnetic setups and the basics of  $\textnormal{T}_n\textnormal{SE}$ are provided in Supplemental Material [12].

. 

\section{\label{sec:level3}Topological Hooke’s law: Ut tensio, sic $\textbf{T}_{\mathrm{st}}$}

Geometric confinement dictates that the axial response of a $\textnormal{T}_n\textnormal{SE}$ depends on the sense of spin rotation. For the counter-clockwise (CCW) in-plane (IP) rotation, an out-of-plane (OOP) spin torque stabilizes contraction (outward) and elongation (inward) [12]; The opposite holds for clockwise (CW) rotation. Fig. 1(d) plots the restoring spin torque $T_{\text{rst}}$ at a $\textnormal{T}_{100}\textnormal{SE}$ boundary against total length $L$,  revealing a robust linear regime up to $\pm20\%$ variation:
\begin{equation}
T_{\text{rst}} =-k\Delta L,  
\end{equation}
with the effective spring constant $k$ negative (positive) for CCW (CW) rotation. The restoring mechanism originates from minute OOP tilting of spins induced by torque imbalance when lacking rigid boundary constraints and consequent precessional torque under strong IP effective field [12]. The deformation enables reversible energy storage. The work done by $T_{\text{rst}}$ during the process is:
\begin{equation}
\mathcal{W}  \approx -\frac{k}{2\lvert \gamma \rvert}(\partial_x \theta|_{\text{bound}}) (\Delta L)^2,  
\end{equation}
where $\theta$  is the spin orientation at TSE boundary and \(\partial_x \theta|_{\text{bound}} =0.0377\,\mathrm{rad/nm}\) [12], consistent with the parabolic energy curve in Fig. 1(d). The energy resides in exchange interaction under compression—in accord with \(E_{\text{exch.den}} =\frac{A}{2}(\partial_x \theta)^2\) [13] and dipole-dipole interaction owing to separation of opposite magnetic charges [12], offering nonvolatile energy storage beyond conventional batteries. The stored potential energy can be released to produce work in a spintronic circuit by spin-motive forces reciprocal to the spin torque [13, 14]. 

The topological nature of SE allows morphological adaptation. In nanostrips of varying width and curvature, the linear "$T_{\text{rst}}$ Vs. $\Delta L$" relation persists (Figs. 1(e) and 1(f)). However, the unit-area $T_{\text{rst}}$ required for a given elongation ratio $\lambda$ decreases with increasing width, indicating a geometry-dependent modulus. In curved strips, a simple Hooke's law no longer suffices: under expansion, stiffness $k$ exhibits a sharp increase, which we attribute to a potential well that traps the walls due to finite curvature [15]. These observations point to a generalized, topology-aware Hooke's law: 
\begin{equation}
\Delta \Gamma  =-T_{\text{rst}}\int_{0}^{\Gamma}\frac{\xi_{r}}{S_{r}\mathrm{E}_r}\, dr,  
\end{equation}
where $\Delta \Gamma$ is the axial change along path $\Gamma$ and $S_{r}$,  $\xi_{r}$, $\mathrm{E}_r$ the cross-sectional area, spin-torque transfer coefficient, and elastic modulus at position $r$. Two remarks are in order: (1) The parameters $S_{r}$, $\xi_{r}$ and $\mathrm{E}_r$ vary with local topological deformation. (2) The path $\Gamma$ can assume an arbitrary curvilinear form. This stands in stark contrast to the conventional elasticity of matter, where the direction of loading cannot be decoupled from the resulting deformation.

\section{\label{sec:level4}Spin elastic theory
}
\begin{figure}
    \centering
    \includegraphics[width=1\linewidth]{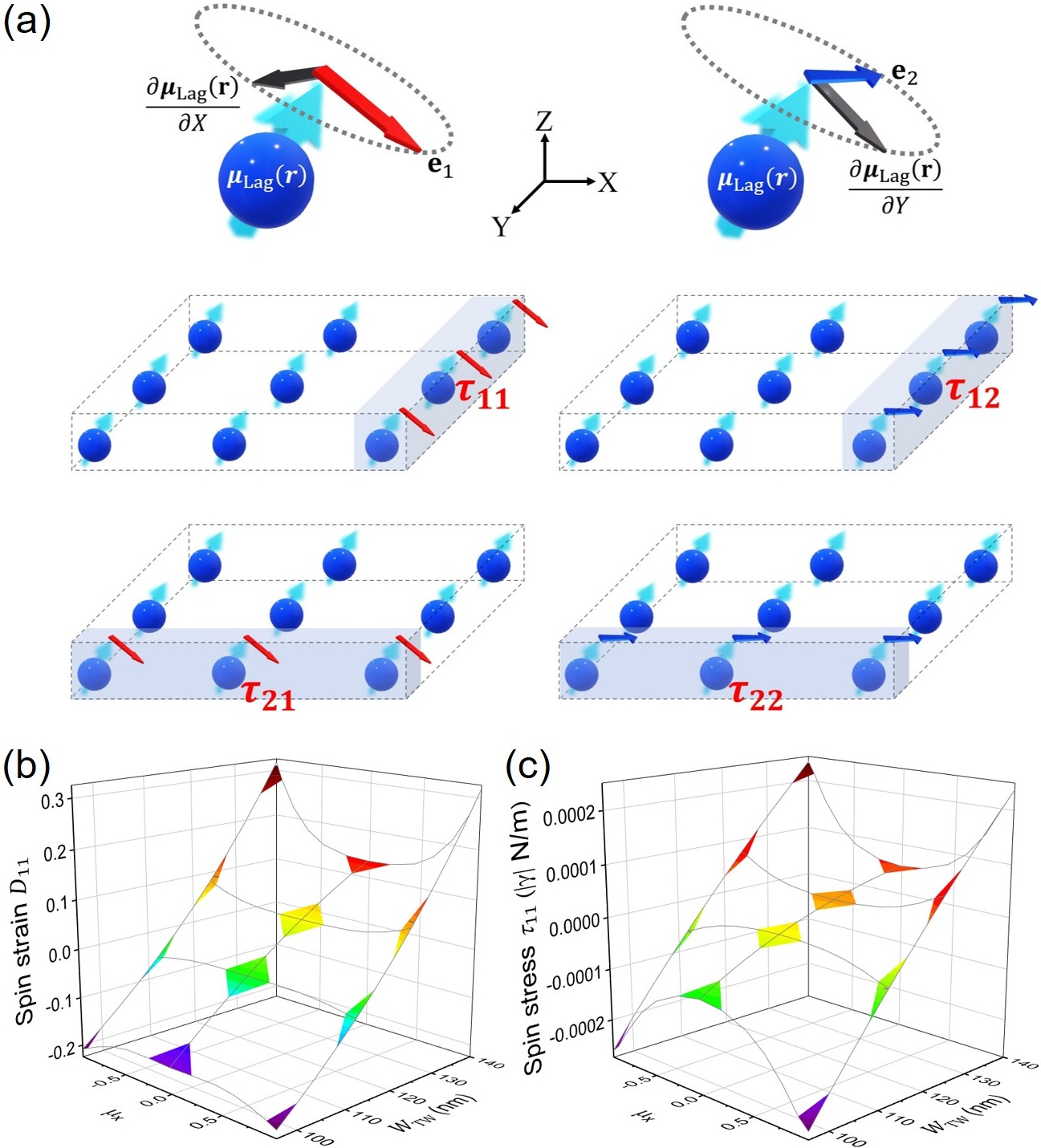}
    \caption{(a) Schematic of spin stress tensor $\tau_{ij}$ where $i$ denotes the normal direction of the surface on which the stress acts, and $j$ the direction of the stress component. The two basis vectors are \(\textbf{e}_1=\bm{\mu}_{\mathrm{Lag}}(\bm{r})\times(\widehat{\frac{\partial \bm{\mu}_{\mathrm{Lag}}(\bm{r})}{\partial X}})\) and \(\textbf{e}_2=\bm{\mu}_{\mathrm{Lag}}(\bm{r})\times(\widehat{\frac{\partial \bm{\mu}_{\mathrm{Lag}}(\bm{r})}{\partial Y}})\)). Spatial distribution (vs. $\mu_x$) and parametric dependence (vs. $W_{\mathrm{TW}}$) of (b) spin strain $D_{11}$ and (c) spin stress $\tau_{11}$ in $\textnormal{T}_n\textnormal{SE}$.}
    \label{fig:placeholder}
\end{figure}

To characterize deformation in SE, we adopt the displacement gradient tensor $\nabla \textbf{u}_{\mu}$—with \(\textbf{u}_{\mu}=x-X \)—as our strain measure, termed spin strain $\textbf{D}$.  Here $X$ and $x$ denote the coordinates of a spin state $\bm{\mu}$ in the initial (Lagrangian) and deformed (Eulerian) configurations. $\nabla \textbf{u}_{\mu}$ (via \(d\textbf{u}_{\mu}=\nabla \textbf{u}_{\mu}d\textbf{X}\)) is isomorphic to the spin-gradient tensor $\nabla \textbf{m}$ (via \(d\textbf{m}=\nabla \textbf{m}d\textbf{r}\)). This mapping holds only for spin textures without connected domains of uniform magnetization. Full spatial coordinatization thus applies to $\leq 2\mathrm{D}$ solitons (skyrmions, merons). 3D textures (Hopfions, skyrmion bundles) admit only partial coordinatization. 

For a fully coordinatizable 2D texture, the spin strain tensor reduces to:
 \begin{subequations}
\begin{equation}
\bm{\textbf{D}}  =\begin{bmatrix}
    \frac{\partial u_1}{\partial X_1} & \frac{\partial u_2}{\partial X_1} \\
    \frac{\partial u_1}{\partial X_2} & \frac{\partial u_2}{\partial X_2}
\end{bmatrix}.  
\end{equation}
In its explicit form, 
\begin{equation}
         \begin{bmatrix}
    D_{11} & D_{12} \\
    D_{21} & D_{22}
\end{bmatrix}=\begin{bmatrix}
    \frac{1}{a_1}-1 & -b_1 \\
    -a_2 & \frac{1}{b_2}-1
\end{bmatrix},
     \end{equation}
 \end{subequations}
with coefficients  $a_i$, $b_i$ defined by the transformation of the Eulerian gradients relative to the Lagrangian configuration: \(\frac{\partial \bm{\mu}_{\text{Euler}}}{\partial i} =a_i \cdot \frac{\partial \bm{\mu}_{\text{Lag}}}{\partial X}+b_i \cdot \frac{\partial \bm{\mu}_{\text{Lag}}}{\partial Y}, i=1,2\). Deformation alters the balance of magnetic interactions, requiring an external spin torque at the boundaries and generating countervailing torques in the interior—a stressed spin state that underpins elastic recovery. To quantify this state, we introduce \textit{spin stress} $\bm{\tau}_{i}$ on a surface $S_i$ (normal along $i$) as the countervailing torque density. Decomposing $\bm{\tau}_{i}$ via \(\bm{\tau}_{i}  =\tau_{i1}\cdot\textbf{e}_1+\tau_{i2}\cdot\textbf{e}_2\) (where \(\textbf{e}_1=\bm{\mu}_{\mathrm{Lag}}(\bm{r})\times(\widehat{\frac{\partial \bm{\mu}_{\mathrm{Lag}}(\bm{r})}{\partial X}})\) and \(\textbf{e}_2=\bm{\mu}_{\mathrm{Lag}}(\bm{r})\times(\widehat{\frac{\partial \bm{\mu}_{\mathrm{Lag}}(\bm{r})}{\partial Y}})\)) gives rise to:
\begin{equation}
\tau_{ij}  =\begin{bmatrix}
    \tau_{11} & \tau_{12} \\
    \tau_{21} & \tau_{22}
\end{bmatrix},  
\end{equation}
which is conjugate to the spin strain tensor $\textbf{D}$. Each component has an independent magnitude and a distinct effect on the volume element (Fig. 2(a)). Evaluation of $\bm{\tau}_{i}(\bm{\mu})$ at each state can be achieved by
\begin{equation}
\bm{\tau}_{i}(\bm{\mu}) =a\lvert \gamma \rvert M_s \cdot \bm{\mu} \times \sum \Delta \textbf{H}_{\mathrm{pas}}(\bm{\mu}),
\end{equation}
where $a$ is the spin lattice constant and $\Delta \textbf{H}_{\mathrm{pas}}(\bm{\mu})$ denotes the adjustments of the passive effective fields (incl. internal exchange field $\textbf{H}_{\mathrm{int.exch}}$, magnetocrystalline anisotropy field $\textbf{H}_{\mathrm{ani}}$, dipolar field $\textbf{H}_{\mathrm{dipo}}$ with respect to self-relaxation. 

We now apply the $\bm{\tau}-\textbf{D}$ framework to quantify the internal state of $\textnormal{T}_{100}\textnormal{SE}$ under different geometric confinements. As in conventional elasticity, the normal strain $\lvert D_{11}\rvert$ increases monotonically with elongation or compression. Its distribution, despite homogeneous medium, is however nonuniform: it grows toward the wall boundaries (Fig. 2(b)) and rises along the transverse direction in central TW [12]. The former reveals the existence of a rigid core and the latter implies that elongation (compression) shifts lower core states upward (downward), producing lateral contraction (expansion)—a Poisson effect of SE [12]. On the other hand, spin stress is not conserved—a fundamental departure from classical mechanics: it varies with position and lacks transferability (Fig. 2(c)), violating Newton's third law. The origin is the nonlocal dipole–dipole interaction: the counter torque at a given site comes from a distributed ensemble of spins, not only from its immediate neighbors. Consequently, torque flow is not fully transmitted. Other interactions (magnetocrystalline anisotropy, Zeeman coupling) act similarly, as part of the counter torque is supplied locally by anisotropy or external fields. In our system, $\tau_{11}$ grows with distance from the TW center, and its spatial derivative follows the same trend, implying a stronger dipolar field near boundaries—consistent with [12]. 

The tensors $\bm{\tau}$ and $\textbf{D}$ encode the full magnetization state. Given the initial $\textbf{D}$ and loading history, the evolution of $\textbf{D}$—and thus of the magnetization—follows uniquely. Beyond the conventional description by spin $\textbf{S}$ and effective field $\textbf{H}_{\mathrm{eff}}$, the $\bm{\tau}-\textbf{D}$ framework offers a novel perspective that renders SE deformation amenable to intuitive visualization and physical insight—absent from energy-minimization-based micromagnetic calculations. Importantly, in local-interaction (exchange, magnetocrystalline anisotropy and Zeeman) dominated systems, there emerges locally bound properties as the spin stress transferability and constitutive relation (e.g., elastic modulus) connecting the two tensors, which reduces micromagnetic calculations to a classical deformation problem. 

\begin{figure*}
    \centering
    \includegraphics[width=0.6\linewidth]{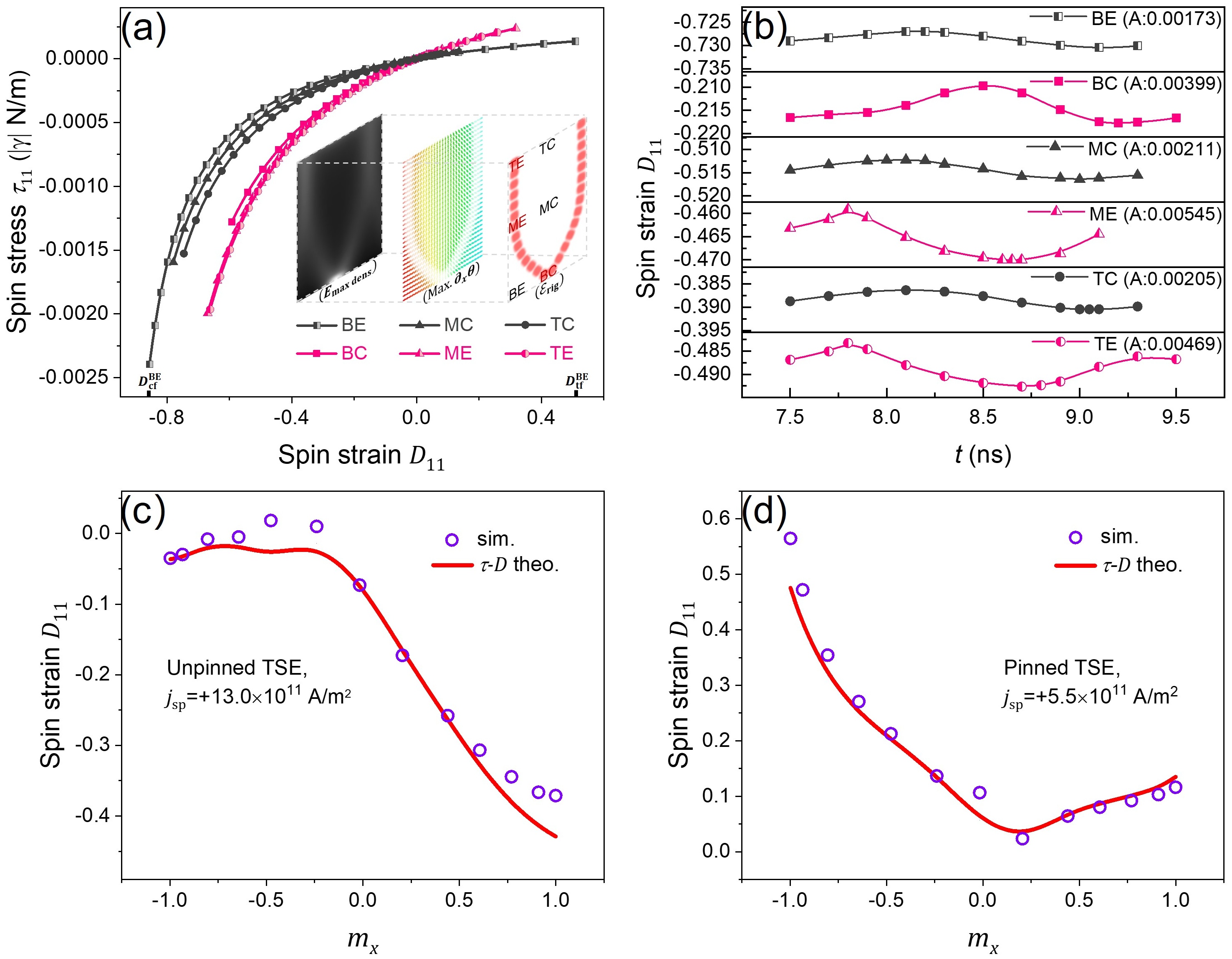}
    \caption{(a) $\tau-D$ curves for six representative spin states in TW: $\mu_x=1$ at TE, ME, BE; and $\mu_x=0$ at TC, MC, BC. $D_{\text{cf}}^{\text{BE}}$ and $D_{\text{tf}}^{\text{BE}}$ mark the critical compressive and tensile failure strains for the BE curve. Inset: spatial correlation of elevated energy density ($E_{\text{max dens}}$), abrupt magnetization transitions (Max. $\partial_x \theta$), and higher rigidity ($\mathcal{E} _{\text{rig}}$). (b) Strain oscillations at the six sites in $\textnormal{T}_{136}\textnormal{SE}$ (\(W_{\mathrm{TW}}=60\,\mathrm{nm}\)) under a spin stress wave, recorded 1770 nm from the source. Waves are excited by an $x-$directional ac magnetic field \(\textbf{H}_{\text{ac}}=h_0\sin(2\pi ft)\hat{\bm{x}}\) (\(f=0.5\,\mathrm{GHz}\), \(h_0=50 \,\mathrm{Oe}\)) applied to a narrow central region (\(\mu_y\in(-1,1)\)) of the $\textnormal{T}_{136}\textnormal{SE}$; \(\alpha=10^{-6}\). (c) Spin strain $D_{11}$ as a function of $m_x$ for an unpinned $\textnormal{T}_{4}\textnormal{SE}$ at spin polarized current \(j_{\text{sp}} =13.0\times 10^{11} \,\mathrm{A/m^2}\), and (d) for a pinned $\textnormal{T}_{4}\textnormal{SE}$ at \(j_{\text{sp}} =5.5\times 10^{11} \,\mathrm{A/m^2}\). Red curves: $\tau-D$ analytics; purple circles: micromagnetic simulations.}
    \label{fig:placeholder}
\end{figure*}

To capture the stress-strain response of the system, we compute $\tau-D$ curves for representative positions (Fig. 3(a)): top edge (TE), middle edge (ME), bottom edge (BE) at $\mu_x=1$; and top center (TC), middle center (MC), bottom center (BC) at $\mu_x=0$. Divergence of the curves reflects site-dependent elastic nature. Beyond the proportionality and failure limits, the \textit{spin modulus}—defined as \(\mathrm{E}_{11}=\frac{\tau_{11}}{D_{11}}\)—at each strain level can be readily extracted from the curves. Fig. 3(c) and 3(d) show quantitative comparison between the $\tau-D$ theory and direct simulations for the strain distribution within TW for an unpinned $\textnormal{T}_{4}\textnormal{SE}$ under large spin polarized current \(j_{\text{sp}} =13.0\times 10^{11} \,\mathrm{A/m^2}\) and a pinned $\textnormal{T}_{4}\textnormal{SE}$ under \(j_{\text{sp}} =5.5\times 10^{11} \,\mathrm{A/m^2}\), respectively. The analytical model employing $\mathrm{E}_{11}$ captures the asymmetric deformation landscape well; the slight deviation mainly originates from coarse assignment of $\mathrm{E}_{11}$. 

The $\tau-D$ curves bifurcate. The red branch (mostly edge sites) corresponds to higher rigidity—the BE-BC exchange originates from limited spin stress transferability across BC due to the enhanced dipolar field there [12]. Identifying such high-rigidity sites can help locate regions of elevated energy density and abrupt magnetization transitions; their correlation is shown in the inset of Fig. 3(a). Counter intuitively, these rigid sites are less resistant to perturbation—opposite to conventional matter. As shown in Fig. 3(b), sites BC, ME and TE facilitate strain-oscillation propagation, forming an effective channel for “spin stress wave” (discussed later). This abnormality may attribute to the strong magnetization nonlinearity bound to these sites as nonlinear magnetization is prerequisite medium for spin stress waves. 

To describe the elastic dynamics of SE, we treat constituent solitons as quasiparticles—focusing on their positions and velocities. A key simplification emerges from the symmetry of topological solitons. Within a typical SE, there exist high-symmetry points $\Lambda_{n}$ at which the dipole-dipole interaction cancels out exactly, yielding a locally fixed constitutive relation and no net contribution to spin stress [12]. In the simplest case involving only exchange interaction (in the continuum limit, exchange field \(\textbf{H}_{\text{exch}}=-\frac{2A}{\mu_{\text{0}}M_{\text{S}}} \nabla ^2 \bm{\mu}\)), the constitutive relation writes:
\begin{equation}
\bm{\tau}=-\frac{2\lvert \gamma \rvert A}{\mu_{\text{0}}}\bm{\mu}\times \bigg(\textbf{D}\begin{bmatrix}
    \frac{\partial \bm{\mu}_{\text{Lag}}}{\partial X_1} & 0 \\
    0 & \frac{\partial \bm{\mu}_{\text{Lag}}}{\partial X_2}
\end{bmatrix}\bigg).  
\end{equation}
Together with the geometric relation Eq. (5a) and the equations of motion at high-symmetry points:
\begin{subequations}
\begin{equation}
(\textbf{v}\cdot \nabla)\bm{\mu}=-\frac{1}{(1+\alpha^2)M_{\text{S}}}(\alpha \bm{\mu}\times \bm{\mathcal{T}}+\bm{\mathcal{T}}),  
\end{equation}
\begin{equation}
\bm{\mathcal{T}}=\nabla \times \bm{\tau}+\bm{\mathcal{T}}_{\text{ext}},  
\end{equation}
\end{subequations}
where \(\textbf{v}=\frac{\partial \textbf{u}}{\partial t}\) is the instantaneous velocity of the spin state at a high-symmetry point, $\bm{\mathcal{T}}$ the volumetric spin torque density, and \(\bm{\mathcal{T}}_{\text{ext}}=p_1\frac{\partial \bm{\mu}_{\text{Lag}}}{\partial X_1}+p_2\frac{\partial \bm{\mu}_{\text{Lag}}}{\partial X_2}\) external contribution, we obtain spin elastodynamic equation: 
\begin{equation}
\begin{aligned}
    \frac{\partial u_i}{\partial t} & = \frac{2\lvert \gamma \rvert A}{(1+\alpha^2)\mu_{0}M_{\text{S}}}[-(\alpha+\cot\theta)\nabla^2 u_i+\frac{1}{\sin \theta}\nabla^2 u_j] \\
    & -\frac{1}{(1+\alpha^2)M_{\text{S}}}\frac{(\sin \theta-\alpha\cos \theta)p_i+\alpha p_j}{\sin \theta}
\end{aligned},
\end{equation}
where $\theta$ is the included angle between $\frac{\partial \bm{\mu}_{\text{Lag}}}{\partial X_1}$ and $\frac{\partial \bm{\mu}_{\text{Lag}}}{\partial X_2}$. The first term on the right represents the system-derived elastic restoration which concerns internal strain gradient, while the second term captures the external driving. The spin elastic equilibrium equation has the simple form: 
\begin{equation}
\nabla \times \bm{\tau}=-\bm{\mathcal{T}}_{\text{ext}},  
\end{equation}
balancing the internal stress divergence with the external torque density. In this regard, the $\bm{\tau}-\textbf{D}$ framework surpasses the Thiele formulism [16,17]—which assumes rigid soliton profiles and neglects the contribution of internal deformations to collective motion—offering a more rigorous description of solitonic dynamics. 

\section{\label{sec:level5}Oscillations and resonance }
A fundamental question in spin elastodynamics is whether elastic oscillations can occur in SE—another defining signature of spin elasticity. In ferromagnets, spin dynamics obeys the first-order LLG equation, as opposed to the second-order dynamics in antiferromagnets [18] or Newtonian mechanics. Consequently, SE possesses no intrinsic inertia and, in principle, should not oscillate spontaneously about equilibrium. Yet, upon releasing an initially compressed $\textnormal{T}_{100}\textnormal{SE}$, clear oscillations emerge—in a sharp triangular waveform seldom seen in nature (Fig. 4(a)). To understand this behavior, we compare the TW profiles at rest and in steady motion. Higher velocity enhances out-of-plane tilting as expected and narrows the TW width [12]. This narrowing produces spin strain and amplifies strain gradient, consistent with Eq. (10) at higher velocities. A persistently maintained strain gradient sustains inertial motion or inertia, rendering elastic oscillations possible. As the Gilbert constant increases, underdamped, critically damped, and overdamped regimes appear sequentially. The critical damping constant is \(\alpha=0.0013\).

\begin{figure*}
    \centering
    \includegraphics[width=0.6\linewidth]{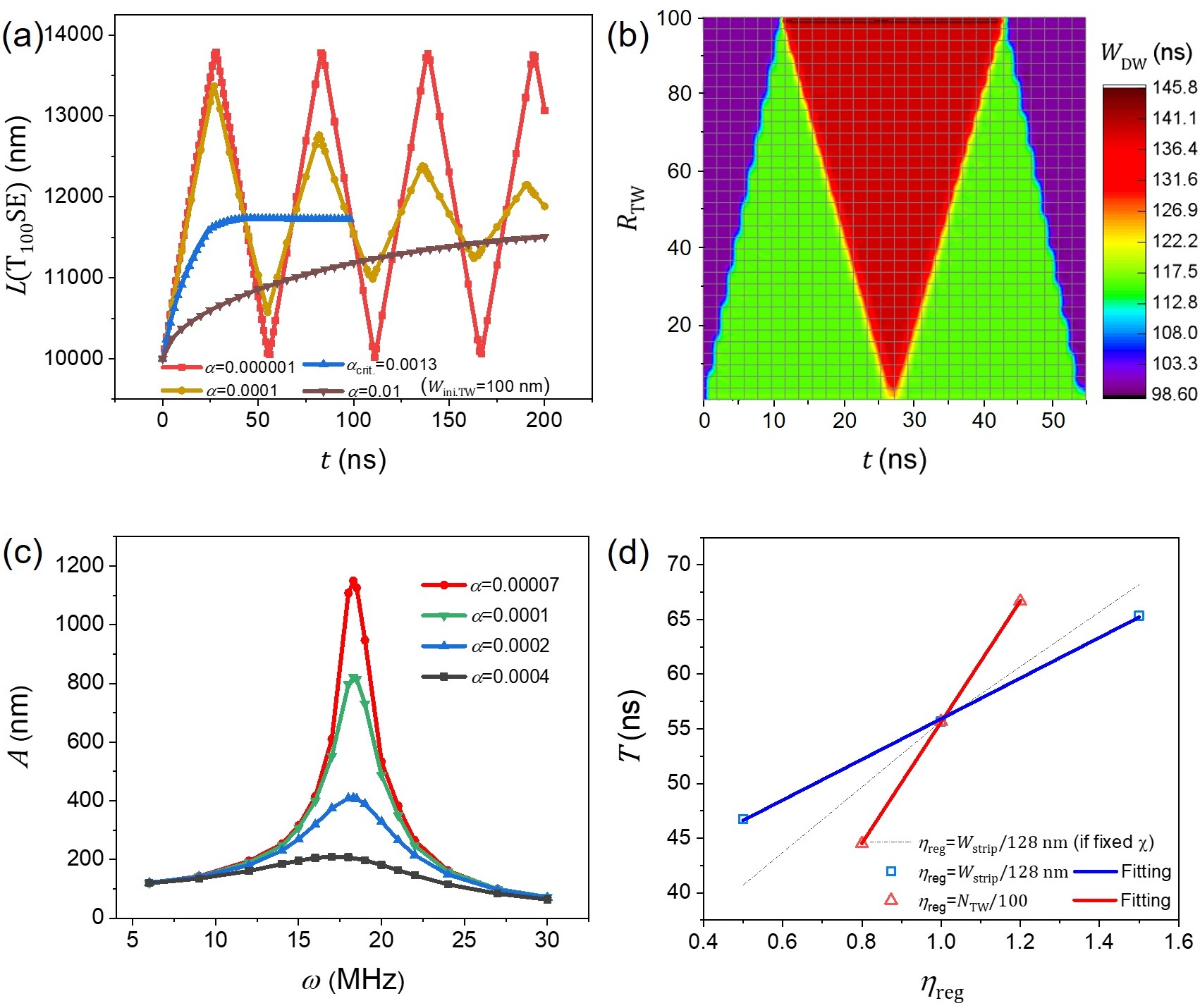}
    \caption{(a) Oscillations of $\textnormal{T}_{100}\textnormal{SE}$ released from compression (\(W_{\mathrm{TW}}=100\,\mathrm{nm}\)) for different damping constants; \(\alpha_{\mathrm{crit}}=0.0013\) is the critical damping constant. (b) Time resolved $W_{\mathrm{TW}}$ distribution during oscillation. (c) Resonance spectra for different damping constants (\(\alpha=0.00007, 0.0001, 0.0002, 0.0004\)) under a sinusoidal spin polarized current with peak amplitude \(2.0\times 10^{11} \,\mathrm{A/m^2}\). (d) Period modulation by normalized strip width ($\frac{W_{\mathrm{strip}}}{W_{\mathrm{strip,def}}}$, with \(W_{\mathrm{strip,def}}=128\,\mathrm{nm}\)) and normalized TW number ($\frac{N_{\mathrm{TW}}}{N_{\mathrm{TW,def}}}$, with \(N_{\mathrm{TW,def}}=128\,\mathrm{nm}\))}
    \label{fig:placeholder}
\end{figure*}

A puzzle remains: Eq. (10) predicts nonlinear oscillations of TSE for synchronized TW deformation, contradicting the observed triangular waveform. Time resolved $W_{\mathrm{TW}}$ distribution manifests serial kinetics (Fig. 4(b)):  initially compressed, morphological‑kinematic state transition ([100 nm, 0 m/s] → [pseudo-equilibrium 115.8 nm, -138.9 m/s] → [138.5 nm, 0 m/s]) occurs sequentially, owing to gradual boundary constraint release and inertial motion; at maximum expansion, elastic recovery and inertia drive sequential contraction ([138.5 nm, 0 m/s] → [115.8 nm, 134.3 m/s] → [100 nm, 0 m/s]), restoring the initial state.

Oscillations with different amplitudes share the same frequency ($\sim 18 \,\mathrm{MHz}$) [12]—a natural frequency. Resonance is confirmed under a sinusoidal spin-polarized current excitation (Fig. 4(c)), with the peak coinciding with the natural frequency. The resonance amplitude grows as damping decreases. The serial kinetics implies a fixed proportionality $\chi$ between the serial velocity and averaged initial spin strain. Fig. 4(d) shows successful period modulation by the number of TWs ($N_{\mathrm{TW}}$) and the strip width ($W_{\mathrm{strip}}$). Width tuning has a weaker effect because the normalized TW width ($\frac{W_{\mathrm{TW}}}{W_{\mathrm{TW,def}}}$) is discounted relative to the normalized strip width ($\frac{W_{\mathrm{strip}}}{W_{\mathrm{strip,def}}}$) [12], and because $\chi$ increases for topologically enlarged TWs.

\section{\label{sec:level6}Spin stress wave}

Elasticity and inertia—the two prerequisites for stress waves—are now established in SE, implying the existence of their dynamic counterpart: \textit{spin stress waves}. To test this prediction, we apply an $x-$directional ac magnetic field to a narrow central region (\(\mu_y\in(-1,1)\)) of a $\textnormal{T}_{136}\textnormal{SE}$. Spin waves are excited and propagate along the assembly, as seen in the $m_y$ profile recorded 1650 nm from the source (Fig. 5(a)). What is unusual, these spin waves carry a companion: a dynamically oscillating $\tau$—a spin stress wave (Fig. 5(b))—at the same location, leading the local spin oscillation by $\pi/4$ in phase. That is, the maximum stressed state corresponds to equilibrium orientation of local spins. Notably, the energy density oscillates at the same frequency as the $\tau$ wave—rather than twice that frequency—and is $\pi/2$ out of phase, a departure from conventional expectations (Fig. 5(c)). This anomaly arises because, in a SE hosting $\tau$ wave (hence spatial expansion and contraction), fluctuations in soliton density become the dominant factor governing energy density [12]. The $\pi/2$ phase shift follows naturally from the fact that peak stress corresponds to maximum TW expansion. Spin stress waves enable the elucidation of nonequilibrium dynamic responses under external loading, thereby completing the theoretical edifice of spin elasticity.

\begin{figure*}
    \centering
    \includegraphics[width=0.9\linewidth]{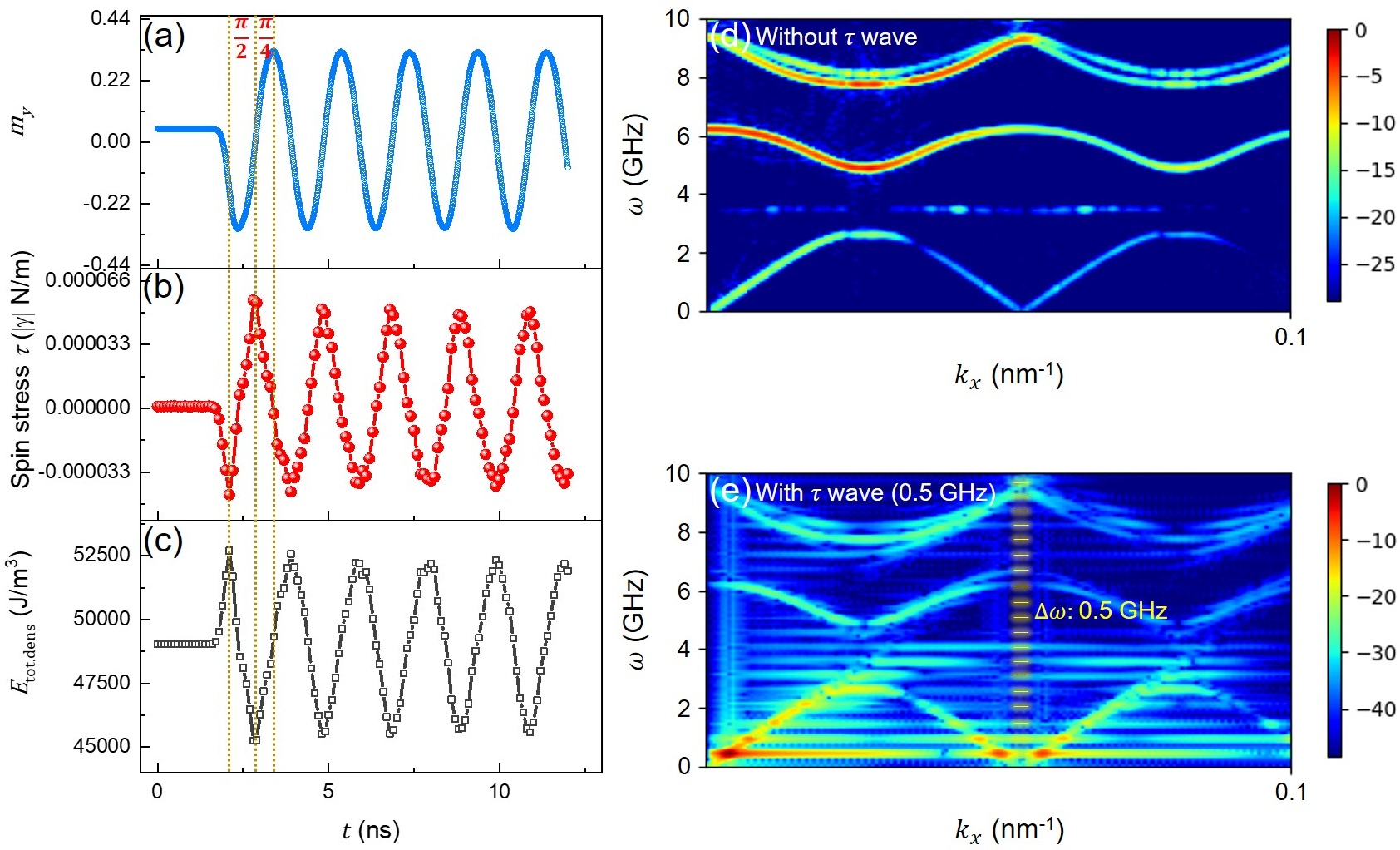}
    \caption{Comparison of (a) spin waves, (b) spin stress waves, and (c) energy density waves in a $\textnormal{T}_{136}\textnormal{SE}$. Waves are excited in the same fashion as in Fig. 3(b). The spin wave is recorded at a fixed spin site 1650 nm from the source. The spin stress wave is computed from \(\tau_{11, \mu_{x}=1}=\bar{\mathrm{E}}_{11, \mu_{x}=1} \cdot D_{11, \mu_{x}=1}\), where $D_{11}$ is the spin strain and \(\bar{\mathrm{E}}_{11, \mu_{x}=1}=0.00098\) the estimated average spin modulus at \(\mu_x=1\) near \(x=1650\,\mathrm{nm}\).}
    \label{fig:placeholder}
\end{figure*}

Conceptually, spin stress waves are a subclass of spin waves. Unlike ordinary spin waves (Fig. 5(d), (e)), they are not gap confined eigenmodes. Their frequency is drive locked and nondispersive, guaranteeing full band transmission. Nonlinearity of the periodically driven spin stress wave spontaneously generates higher harmonics, producing a series of equally spaced flat bands—a nondispersive magnon frequency comb (MFC). Spin stress waves can coexist and hybridize with ordinary spin waves, enabling dual mode transmission.

\section{\label{sec:level7}Conclusions }

We have introduced and established \textit{spin elasticity}—a paradigm unveiling that elasticity, long regarded as exclusive to ordinary matter, also operates in the spin degree of freedom. The key findings—a large-range topological Hooke’s law, spontaneous oscillations and resonance, and spin stress waves—evidence a new elastic world. By developing a continuum $\bm{\tau}-\textbf{D}$ theory, we unify the elastic behaviors of matter and spin within a single theoretical framework. This work fills the missing spin chapter in the elastic picture and opens a distinct frontier for spintronics, where deformation, rather than magnetization alone, becomes a key variable—paving the way for \textit{spin-elastronic} devices with unprecedented functionality, as demonstrated by a series of proof-of-concept applications in the Supplemental Material [12]. 

\vspace{2em}

\begin{acknowledgments}
This work was supported by National Key R\&D Program of China (Grants No. 2025YFA1411302 and No. 2022YFA1402802), the National Natural Science Foundation of China (NSFC) (Grants No. 12374103, No. 12434003, No. 12134017, No. 12574131, No. 11974250 and No. U2541261), and Sichuan Science and Technology Program (Grant No. 2025NSFJQ0045). Z. G. conceived the research, developed the theory, performed the numerical calculations, and wrote the paper. T. Z., F. W., J. H., P. Y. and X. H. commented on the manuscript. All authors read and approved the final manuscript. 
\end{acknowledgments}

\appendix

\counterwithin{figure}{section}

\renewcommand{\figurename}{FIG.}

\nocite{*}

\bibliography{apssamp}

\end{document}